\documentclass[twocolumn,showpacs,aps,prl,superscriptaddress]{revtex4}

\usepackage{graphicx}
\usepackage{dcolumn}
\usepackage{amsmath}
\usepackage{epsfig}

\input{pubboard/babarsym}

\newcommand{\BABARPubNumber}  {07/044}

\newcommand{\SLACPubNumber} {13000}
\newcommand{\LANLNumber} {0711.2047}

\def\figurebox#1#2#3{
    \def\arg{#3}
    \ifx\arg\empty
    {\hfill\vbox{\hsize#2\hrule\hbox to #2{\vrule\hfill\vbox to #1{\hsize#2\vfill}\vrule}\hrule}\hfill}
    \else
    {\hfill\epsfbox{#3}\hfill}
    \fi}

\long\def\inst#1{\par\nobreak\kern 4pt\nobreak
    {\it #1}\par\vskip 10pt plus 3pt minus 3pt}

\begin{document}

\preprint{\babar-PUB-\BABARPubNumber} 
\preprint{SLAC-PUB-\SLACPubNumber} 

\begin{flushleft}
\babar-PUB-\BABARPubNumber\\
SLAC-PUB-\SLACPubNumber\\
hep-ex/\LANLNumber\\ [10mm]

\end{flushleft}
\title{ {\large {\bf \boldmath Observation of ${Y(3940) \rightarrow
J/\psi\omega}$ in ${B \rightarrow {\bf{J/\psi}} \omega K}$ at \babar }
} }

%
\author{B.~Aubert}
\author{M.~Bona}
\author{D.~Boutigny}
\author{Y.~Karyotakis}
\author{J.~P.~Lees}
\author{V.~Poireau}
\author{X.~Prudent}
\author{V.~Tisserand}
\author{A.~Zghiche}
\affiliation{Laboratoire de Physique des Particules, IN2P3/CNRS et Universit\'e de Savoie, F-74941 Annecy-Le-Vieux, France }
\author{J.~Garra~Tico}
\author{E.~Grauges}
\affiliation{Universitat de Barcelona, Facultat de Fisica, Departament ECM, E-08028 Barcelona, Spain }
% ...
\collaboration{The \babar\ Collaboration}
\noaffiliation

\date{\today}
\begin{abstract}
We present a study of the decays $B^{0,+}\rightarrow J/\psi\omega
K^{0,+}$ using $383\times 10^{6}$ $B\bar{B}$ events obtained with the
\babar\ detector at PEP-II. We observe $Y(3940) \rightarrow J/\psi
\omega$, with mass $3914.6 ^{+3.8}_{-3.4} (stat) \pm{2.0} (syst)$
\mevcc\/, and width $34^{+12}_{-8}(stat)\pm{5}(syst)$ \mev\/. The
ratio of $B^0$ and $B^+$ decay to $YK$ is
$0.27^{+0.28}_{-0.23}(stat)^{+0.04}_{-0.01}(syst)$, and the relevant
$B^0$ and $B^+$ branching fractions are reported.
\end{abstract}

\maketitle

The BELLE Collaboration has reported evidence for the
$X(3940)$~\cite{Abe:2007jn}, the $Y(3940)$~\cite{Abe:2004zs}, and the
$Z(3930)$~\cite{Uehara:2005qd}. The mass and width values are the same
within error, the states have positive $C$ parity, and spin-parity
$(J^P)$ $2^+$ is favored for the $Z$, which may then be the first
radial excitation of the $\chi_{c2}(3556)$, i.e., a charmonium
state. The mass and width consistency with the $X$ and $Y$ suggests
the possibility that these may be the $Z$ in different production
contexts. The $Z$ was found in two-photon production of $D\bar{D}$, so
that it may be a charmonium state. The $X$ was observed in
$e^+e^-\rightarrow J/\psi X$, and decays mainly to $D^{\ast}\bar{D}$,
suggesting a charmonium interpretation. In contrast, the $Y$ was found
in $B\rightarrow YK$, $Y\rightarrow J/\psi\omega$, which is
OZI suppressed for a charmonium state~\cite{Eichten:2007qx}. Also, an
analysis of $B\rightarrow K D\bar{D}$ and $B\rightarrow K
D^{\ast}\bar{D}$~\cite{:2007rva} shows no evidence for the $Y$ (nor
for the $X$ or $Z$), although $\psi(3770)\rightarrow D\bar{D}$ and
$X(3872)\rightarrow D^{\ast}\bar{D}$ are observed. Other possibilities
for the nature of this state, already suggested for the $X(3872)$,
include a hybrid charmonium-gluon bound state, $c\bar{c}g$
~\cite{Close:2005iz,Close:2007ny}, a molecular state of a
$c\bar{c}(u\bar{u}+d\bar{d})$
system~\cite{Tornqvist:2004qy,Braaten:2003he,Swanson:2003tb,Voloshin:2004mh,Voloshin:2007hh},
or a multiquark state~\cite{Maiani:2004vq}. The $S-$wave molecule
model~\cite{Braaten:2003he} predicts a very small $B^0/B^+$ ratio for
$B\rightarrow KX(3872)$.  The previous low value has been
confirmed~\cite{Aubert:2008gu}, although the uncertainties are still
large, so that a measurement of this ratio for the $Y(3940)$ may be
important to an understanding of this state.

In this Letter, we examine the decays $B^{0,+}\rightarrow J/\psi \pi
^{+}\pi ^{-}\pi ^{0}K^{0,+}$~\cite{chargeconj}, with $\pi^+\pi^-\pi^0$
mass in the $\omega$ region. We confirm the $Y(3940)$, improve the
precision of the mass and width significantly, and measure the
$(B^0/B^+)$ production ratio for the first time. Branching fraction
values for $B\rightarrow YK$, $Y\rightarrow J/\psi\omega$, and for
$B\rightarrow J/\psi \omega K$ are obtained for $B^0$ and $B^+$ decay
separately; each is a first measurement.

The data were collected with the {\babar}
detector~\cite{Aubert:2001tu} at the PEP-II asymmetric-energy
$e^{+}e^{-}$ storage rings operating at the $\FourS$ resonance. The
integrated luminosity for this analysis is 348 fb$^{-1}$. The decays
$B^{0,+}\rightarrow J/\psi \pi ^{+}\pi ^{-}\pi ^{0}K^{0,+}$ are
reconstructed as follows (Table~\ref{table-mass}). A candidate $J/\psi
\rightarrow e^{+}e^{-}$ $(\mu^+\mu^-)$ decay has invariant mass in the
$J/\psi$ mass region, and is then constrained to the nominal
mass~\cite{Yao:2006px}. A $K_S^{0}$ candidate has $\pi^+\pi^-$
invariant mass in the $K_S^0$ region. The $J/\psi$ and $K_S^0$
distributions from the $B$ signal region show no significant
background. A $\pi^0$ candidate consists of a photon pair with
invariant mass in the $\pi^0$ region. After a $\pi^0$ mass constraint,
an $\omega \rightarrow \pi ^{+}\pi ^{-}\pi ^{0}$\ candidate has
invariant mass in the $\omega$ region. We form a $B^{+}\left(
B^{0}\right) $ candidate by combining $J/\psi $, $\omega $ and
$K^{+}$~\cite{Aubert:2002rg} $(K_S^0)$ candidates.

We define the $B$ signal region using the center of mass (c.m.) energy
difference $\Delta E\equiv E_{B}^{\ast }-\sqrt{s}/2$, and the
beam-energy substituted mass $m_{ES}\equiv \sqrt{\left((
s/2+\vec{p}_i\cdot
\vec{p}_B\right)/E_{i})^2-\vec{p}_B^{~2}}$~\cite{Aubert:2001tu}, where
$\left( E_{i},\vec{p}_i\right)$ is the initial state four-momentum
vector in the laboratory frame (l.f.); $\sqrt{s}$ is the c.m. energy,
$E_{B}^{\ast}$ is the $B$ meson energy in the c.m., and $\vec{p}_B$ is
its l.f. momentum. Signal events have $\Delta E$ $\sim$ zero and
$m_{ES}$ $\sim m_B$; $12\%$ of the events have multiple candidates,
and for these the combination with the smallest $|\Delta E|$ is
chosen.

The selection criteria were established by optimizing
signal-to-background ratio using Monte Carlo (MC) simulated signal
events, $B\rightarrow YK,Y\rightarrow J/\psi \omega $, and background
$B\bar{B}$ and $e^+e^-\rightarrow q\bar{q}$ ($q=u,d,s,c$) events.
\begin{table}[!htb]
\caption{Principal criteria used to select $B$ candidates.}
\begin{center}
\begin{tabular}{lrcccl}
\hline\hline
Selection Category & Criterion \\
\hline
$J/\psi \rightarrow \mu ^{+}\mu ^{-}$ mass ($\gevcc$)  & $3.06<m_{\mu \mu }<3.14$ &  \\ 
$J/\psi \rightarrow e^{+}e^{-}$ mass ($\gevcc$)& $2.95<m_{ee}<3.14$ &  \\ 
$K_{S}$ mass ($\gevcc$) & $0.472<m_{\pi\pi}<0.522$ &  \\ 
$\pi ^{0}$ mass ($\gevcc$) & $0.115<m_{\gamma \gamma }<0.150$ &  \\ 
$\omega $ signal region ($\gevcc$) ($B^+$)& $0.7695<m_{3\pi }<0.7965$ &  \\ 
$\omega $ signal region ($\gevcc$) ($B^0$)& $0.7605<m_{3\pi }<0.8055$ &  \\ 
$\DeltaE$  ($\gev$)   ($B^+$)            & $\left | \DeltaE\right|<0.020$ \\
$\DeltaE$  ($\gev$)   ($B^0$)             & $\left | \DeltaE\right|<0.015$ \\
$m_{ES}$ ($\gevcc$)                       & $5.274<m_{ES}<5.284$ \\
B helicity angle $\theta_B$ & $\left| \cos \theta _{B}\right| <0.9$ &  \\ 
Photon helicity angle $\theta_\gamma$& $ \cos \theta _{\gamma } <0.95$ &  \\ 
$\psi \left( 2S\right) $ veto ($\gevcc$) & $3.661<M_{J/\psi \pi \pi}<3.711$ &  \\
\hline
\end{tabular}
\end{center}
\label{table-mass}
\end{table}

The $\cos\theta_B$ distribution ($\theta_B$ is the c.m. polar angle of
the $B$) is proportional to $\sin ^{2}\theta_{B}$; since
$e^+e^-\rightarrow q\bar{q}$ events peak toward $\pm 1$, we require
$|\cos\theta_B|<0.9$. The variable $\cos\theta_{\gamma}$ is the
normalized dot product between the higher momentum photon in the
$\pi^0$ rest frame (r.f.) and the l.f. direction of the $\pi^0$. For
$\pi^0$ decay this distribution is flat; background peaks at $1$,
hence we require $\cos\theta_\gamma<0.95$. Events from $ B\rightarrow
\psi\left( 2S\right) K\pi^0 ,\psi \left( 2S\right) \rightarrow \pi^+
\pi^- J/\psi $, are removed by the $\psi \left( 2S\right) $ veto.

\begin{figure}[!htb]
\begin{center}
\includegraphics[width=8.4cm]{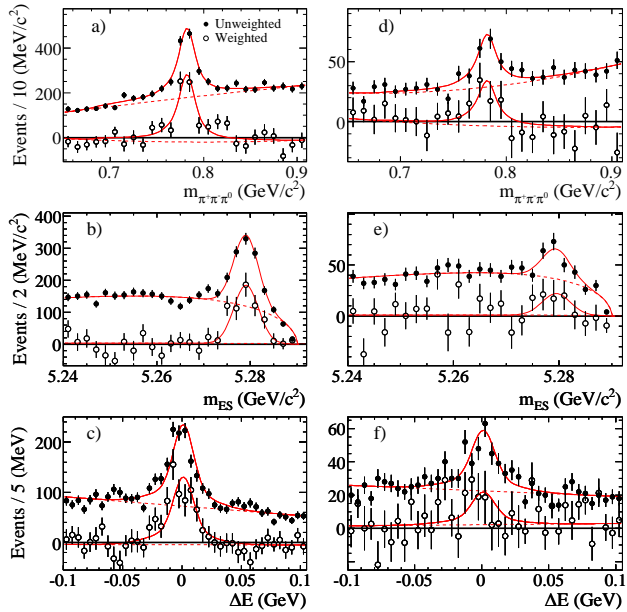}
\caption{(a)-(c) ((d)-(f)) The $3\pi$ mass, $\mes$, and $\Delta E$
distributions for the $B^+$ ($B^0$) mode; solid (open) dots are for
unweighted (weighted) events. The solid (dashed) curves represent
signal plus background (background).}
\label{fig:jpkub}
\end{center}
\end{figure}

The $3\pi $ mass, $m_{ES}$, and $\Delta E$ distributions are shown in
Fig.~\ref{fig:jpkub}, where we apply all Table~\ref{table-mass}
criteria except the requirement on the variable plotted. We fit the
$3\pi$ mass distributions with an $\omega$-meson Breit-Wigner (BW)
line shape (nominal $\omega$\ mass and width~\cite{Yao:2006px})
convolved with a MC-determined triple-Gaussian resolution function as
signal, and a quadratic background function. We fit the $\m_{ES}$
distributions with a signal Gaussian with mass and width fixed from
MC, and an ARGUS background function~\cite{Albrecht:1990cs}, and fit
the $\DeltaE$ distributions with a double-Gaussian signal function
determined from MC, and a linear background function.

There is a large $\omega$ signal for the $B^+$ mode, and a smaller
signal for $B^0$; the $m_{ES}$ and $\DeltaE$ distributions exhibit
clear $B$ signals. We establish the correlation between the $\omega$
and $B$ signals with a projection procedure based on the $\omega$
decay angular distribution. The helicity angle, $\theta_h$, is the
angle between the $\pi^+$\ and $\pi^0$\ directions in the $\pi^+\pi^-$
r.f.. The $\cos\theta_h$ distribution is proportional to
$\sin^2\theta_h$, and the $\omega$ signal is projected by giving the
$i^{th}$ event weight $w_i=\frac{5}{2}(1-3\cos^2\theta_h^i)$. The
effect is shown in Fig~\ref{fig:jpkub}. For the $B^+$ mode, the omega
signal survives, and background is removed.  For the $B^0$ mode the
effect is qualitatively similar. Confirmation is obtained from a fit
to the $3\pi$ mass distribution in each interval. We conclude that
there is one-to-one correspondence between the $\omega$ and $B$-meson
signals in $m_{ES}$ and $\DeltaE$, and that, at the present level of
statistics, the $3\pi$ system in the $\omega$ mass region results
entirely from $\omega$ decay for $B\rightarrow
J/\psi\pi^+\pi^-\pi^0K$. The $\omega-m_{ES}$ (or $\DeltaE$) signal
correlation is important to an analysis of the $J/\psi\omega$
threshold mass region. Near threshold, the $3\pi$ mass distribution
above the $\omega$ mass is limited in range and distorted in
shape. The $m_{ES}$ distribution is not affected, and so we use
$m_{ES}$ fits to extract the $J/\psi\omega$ mass distribution.

\begin{figure}[!htb]
\begin{center}
\includegraphics[width=8.4cm]{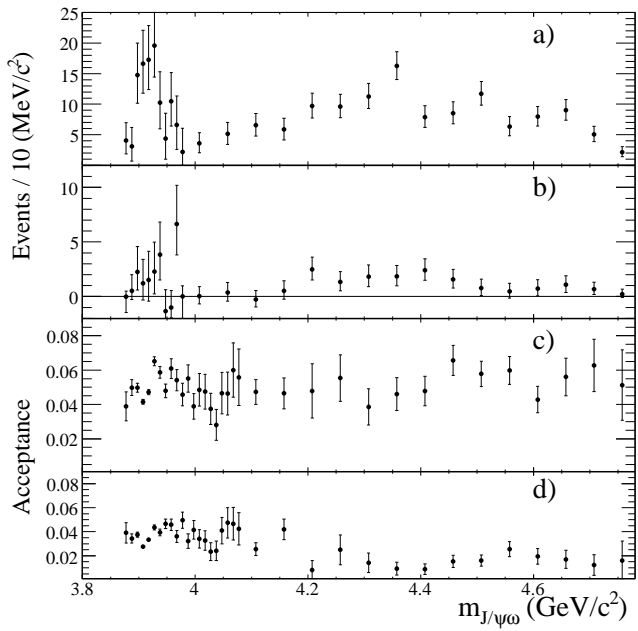}
\caption{The $J/\psi\omega$ mass distribution from the $m_{ES}$ fits
 for (a) $B^+$, and (b) $B^0$ decay. The acceptance as a function of
 $J/\psi\omega$ mass (c) for the $B^+$, and (d) for the $B^0$ mode.}
\label{fig:2}
\end{center}
\end{figure}

For each $B$ decay mode, the $m_{ES}$ distribution in each interval of
$J/\psi\pi^+\pi^-\pi^0$ invariant mass is fitted to extract the
$J/\psi\omega$ signal. The $m_{ES}$ signal, and ARGUS background,
functions are those of Fig.~\ref{fig:jpkub}; the fits use a binned
Poisson likelihood function with signal and background normalizations
free~\cite{James:1975dr}. All fits converge properly and provide good
descriptions of the data. From threshold to 4 \gevcc\/, the
$J/\psi\omega$ mass resolution varies from $5-8$ \mevcc\/, and so
in this region the spectrum is investigated in 11 intervals of
10\mevcc\ starting at $3.8725$ \gevcc\/. At higher mass, there is no
evidence of narrow structure, and we show the results in $50$ \mevcc\
intervals. In Fig.~\ref{fig:2}, there is a clear enhancement near
threshold for $B^+$ decay, while at higher mass no structure is
apparent. The total $B^+$ ($B^0$) signal in Fig.~\ref{fig:2} is
$236^{+18}_{-15}$ ($32^{+8}_{-7}$) events of which $109^{+15}_{-13}$
($16^{+7}_{-6}$) have $J/\psi\omega$ mass less than 4 \gevcc\/
(statistical errors only).

We correct the mass distributions of Fig.~\ref{fig:2} for efficiency
and resolution. In the MC simulation of the $Y$\ signal, we assume
phase space decays of $B\rightarrow YK$ and $Y\rightarrow J/\psi
\omega$, but use the correct angular distribution for $\omega$\
decay. Initially we used a relativistic $S$-wave BW line shape with
$M(Y)=3.940$ \gevcc and $\Gamma(Y)=0.06$ GeV~\cite{Abe:2004zs}. Mass
resolution effects result in a net flow of events away from the peak
mass value. For a given mass interval we define acceptance as the
ratio of events reconstructed in that interval to events generated in
the interval; this accounts for efficiency and resolution effects. The
acceptance-corrected spectrum is fit to a relativistic BW line shape
without convolving resolution, since the acceptance correction takes
this into account. We obtain values of $M(Y)$ and $\Gamma(Y)$ which
are smaller than in the initial simulation, and so generate new MC
samples with the new values in order to correctly reproduce resolution
effects. This iterative procedure converges quickly, and the
acceptance results in Figs.~\ref{fig:2}(c), (d) are obtained with
$M(Y)=3.915$ \gevcc and $\Gamma(Y)=0.02$ GeV. The dip at $\sim 3.91$
\gevcc\/ is due to net flow of events away from the resonance maximum
because of mass resolution. At lower mass, the acceptance is slightly
lower than at higher mass because of the proximity to
threshold. Although the acceptance variation in the $Y$ signal region
is significant, the effect on the $Y$ fit parameters, and on the
corrected number of signal events, is small because of the large
statistical uncertainties on the data.

\begin{figure}[!htb]
\begin{center}
\includegraphics[width=8.4cm]{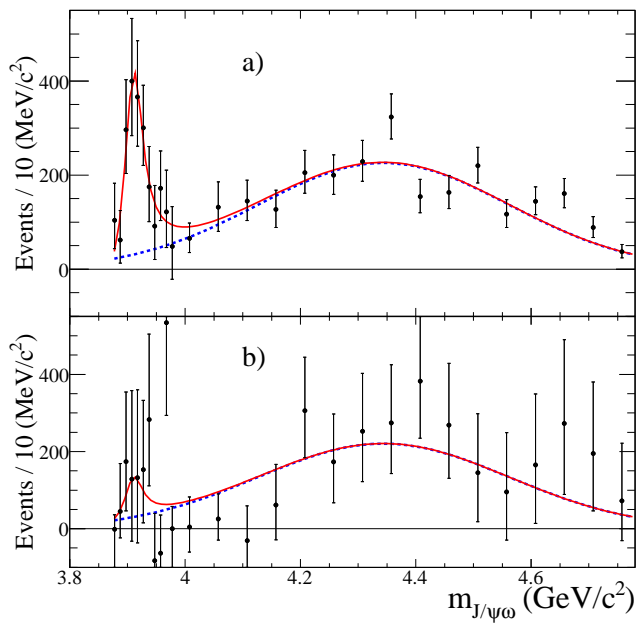}
\caption{The corrected $J/\psi\omega$ mass distribution for (a) $B^+$
 and (b) $B^0$ decay. Each solid (dashed) curve represents the total
 fit function (the nonresonant contribution).}
\label{fig:3}
\end{center}
\end{figure}

The decrease in acceptance at high mass in Fig.~\ref{fig:2}(d) results
from decreasing $K_S^0$ l.f. momentum. The decay pion reconstruction
probability decreases because its l.f. momentum is too small, or
because the decay opening angle is so large that the pion does not
intersect enough detector planes. Fig.~\ref{fig:3} shows the corrected
mass distributions. Below $\sim 4$ \gevcc\ we correct
interval-by-interval, while for higher mass we use a linear fit to the
$J/\psi\omega$ mass dependence. The $B^0$ data are corrected for
$K_L^0$ and $K_S^0\rightarrow \pi^0\pi^0$ decays.

We associate the near-threshold enhancement in Fig.~\ref{fig:3}(a)
with $Y$ production~\cite{Abe:2004zs}, and obtain the mass, width and
decay rate from $\chi^2$ fits. The fit function consists of a
relativistic $S-$wave BW describing the $Y$ and a Gaussian nonresonant
contribution. The corrected $B^+$ and $B^0$ distributions are fitted
simultaneously, with mass, width and Gaussian parameters as common
free parameters. The fit describes the data well ($\chi^2/NDF =
45/44$, NDF$=$number of degrees of freedom ), as shown in
Fig.~\ref{fig:3}. In Fig.~\ref{fig:3}(a), the acceptance-corrected
number of events with $J/\psi\omega$ mass less than $3.98$ \gevcc\ is
$2140\pm290(stat)$, while for the Gaussian it is $420\pm
90(stat)$. Our average efficiency of $\sim 5\%$ implies that a
background fluctuation of $\sim 19$ standard deviations would be
required to describe the near-threshold enhancement. This occurrence
has negligible probability, and so we have instead a clear observation
of the $Y(3940)$. The simultaneous fit yields a $Y$ signal of
$1980^{+396}_{-379}(stat)$ events ({\it{i.e.}} magnitude 5.2 standard
deviations) for $B^+$, and $527^{+534}_{-454}(stat)$ for $B^0$.

Since the acceptance-correction procedure may depend on the input MC
$Y(3940)$ line shape, we combine the first 11 mass intervals for data
and MC and make an overall efficiency correction. The results differ
by $1.9\%$, and we incorporate this as a systematic error associated
with the MC line shape. Other systematic errors are estimated by
repeating the entire process, separately varying by $\pm1\sigma$ the
signal peak and width, and the ARGUS parameter, for the $m_{ES}$
fits. The largest systematic uncertainty contributions to the $B^+$
branching fraction are $5-6\%$ due to the uncertainties in the
secondary branching fractions, tracking efficiency, and particle
identification. For $B^0$, the largest contribution is $10\%$ due to
$m_{ES}$ mass variation; secondary branching fractions, particle
identification, tracking and $K_S$ reconstruction efficiency
contribute also. For both modes, there are uncertainties associated
with the number of $B\bar{B}$ events produced, and with MC sample
size. The product branching fraction for $B^+\rightarrow YK^+$,
$Y\rightarrow J/\psi\omega$ is $(4.9^{+1.0}_{-0.9}(stat)\pm
0.5(syst))\times 10^{-5}$, and that for $B^0\rightarrow YK^0$,
$Y\rightarrow J/\psi\omega$ is $(1.3^{+1.3}_{-1.1}(stat)\pm
0.2(syst))\times 10^{-5}$, with upper limit ($95\%$ C.L.) $3.9\times
10^{-5}$ for the latter. The corresponding branching fractions for
$B\rightarrow J/\psi\omega K$ are $(3.5\pm 0.2(stat)\pm
0.4(syst))\times 10^{-4}$, and $(3.1\pm 0.6(stat)\pm 0.3(syst))\times
10^{-4}$, respectively.

We define $R_Y$ and $R_{NR}$ as the ratios between the number of $B^0$
and $B^+$ events (after all corrections) for the $Y$ signal and for
the nonresonant contribution, respectively. Simultaneous fits to
Figs.~\ref{fig:3}(a),(b) yield the values
$R_Y=0.27^{+0.28}_{-0.23}(stat)^{+0.04}_{-0.01}(syst)$ and
$R_{NR}=0.97^{+0.23}_{-0.22}(stat)^{+0.03}_{-0.02}(syst)$; the upper
limit ($95\%$ C.L.) on $R_Y$ is $0.75$. Although the uncertainty is
large, the central value of $R_Y$ is smaller than expected from
isospin conservation. In comparison, $R$ is $0.865 \pm 0.044$ for
$B\rightarrow J/\psi K$~\cite{Yao:2006px} and $0.81 \pm
0.05(stat)\pm0.01(syst)$ for $B\rightarrow\psi(2S)
K$~\cite{Aubert:2008gu}.

The $Y$ mass and width measurements are subject to additional
systematic effects. When MC-generated signal events are fitted using
the input line shape with mass and width as free parameters, the fitted
value of the mass is 1.6 \mevcc lower than the input value of 3.915
\gevcc\/. This results from the limited $3\pi$ phase space near
$J/\psi \omega$ threshold, and so we increase the fitted $Y$ mass
value by 1.6 \mevcc, and assign this as a systematic
uncertainty. Also, we have used an $S$-wave BW line shape to describe
the $Y$. We repeat the fit using a $P$-wave line shape. The fitted mass
value decreases by $1 \mevcc$, and the width increases by $5 \mev$. We
assign these as systematic uncertainties due to the choice of orbital
angular momentum. Finally, a fit to the uncorrected distributions
(Fig.~\ref{fig:2}) yields a mass value 1.4 \mevcc\/ larger, and a
width 4 \mev\ larger, than obtained for the corrected
distributions. The mass dependence of the acceptance depends on the MC
line shape and so systematic uncertainties of 0.7\mevcc\/ and 2 \mev\/,
respectively are associated with the MC line shape choice. These
contributions dominate all other sources of systematic uncertainty,
and the final mass and width values are $(3914.6^{+3.8}_{-3.4}(stat)\pm
2.0(syst))$ \mevcc\/, and $(34^{+12}_{-8}(stat)\pm 5(syst))$ \mev\/,
respectively.

In summary, in the decays $B^{0,+}\rightarrow J/\psi\omega K^{0,+}$ we
find a $J/\psi\omega$ mass enhancement at $\sim 3.915$ \gevcc\/,
confirming the BELLE result~\cite{Abe:2004zs}, but obtain lower mass,
smaller width, and reduce the uncertainty on each by a factor $\sim
3$. The mass is two standard deviations lower than the $Z(3930)$ mass,
and three standard deviations lower than for the $X(3940)$; the width
agrees with the $Z(3930)$ and $X(3940)$ values. The ratio of $B^0$ and
$B^+$ decay to $YK$, $R_Y$, is measured for the first time and found
to be $\sim 3$ standard deviations below the isospin expectation, but
agrees with that for the X(3872)~\cite{Aubert:2008gu}. The ratio for
the nonresonant contribution $R_{NR}$ agrees with the isospin
expectation. We have obtained first measurements of the branching
fractions for $B\rightarrow J/\psi\omega K$ and for $B\rightarrow YK,
Y\rightarrow J/\psi\omega$, for $B^0$ and $B^+$ decays separately.

We are grateful for the excellent luminosity and machine conditions
provided by our \pep2\ colleagues, 
and for the substantial dedicated effort from
the computing organizations that support \babar.
The collaborating institutions wish to thank 
SLAC for its support and kind hospitality. 
This work is supported by
DOE
and NSF (USA),
NSERC (Canada),
CEA and
CNRS-IN2P3
(France),
BMBF and DFG
(Germany),
INFN (Italy),
FOM (The Netherlands),
NFR (Norway),
MIST (Russia),
MEC (Spain), and
STFC (United Kingdom). 
Individuals have received support from the
Marie Curie EIF (European Union) and
the A.~P.~Sloan Foundation.


\begin{thebibliography}{10}
\bibitem{chargeconj}
  {The use of charge conjugate reactions is implied throughout this Letter.}
\end{thebibliography}
\end{document}